\documentclass{article}

\usepackage{PRIMEarxiv}

\usepackage[utf8]{inputenc} 
\usepackage[T1]{fontenc}    
\usepackage{hyperref}       
\usepackage{url}            
\usepackage{booktabs}       
\usepackage{amsfonts}       
\usepackage{nicefrac}       
\usepackage{microtype}      
\usepackage{lipsum}
\usepackage{fancyhdr}       
\usepackage{graphicx}       
\graphicspath{{media/}}     
\usepackage[ruled,vlined]{algorithm2e}
\usepackage{caption}
\usepackage{amsmath}
\usepackage{float}

\title{Fan-Beam CT Reconstruction for Unaligned Sparse-View X-ray Baggage Dataset
}

\author{
  Shin Kim\\
   Independent Researcher  \\
 indepth1024@gmail.com \\ 
  \\
  \url{https://shinkim148.github.io/FBCT-project-page/} \\
\\
}

\begin{document}
\maketitle

\begin{abstract}

\hspace{1em} Computed Tomography (CT) is a technology that reconstructs cross-sectional images using X-ray images taken from multiple directions. In CT, hundreds of X-ray images acquired as the X-ray source and detector rotate around a central axis, are used for precise reconstruction. In security baggage inspection, X-ray imaging is also widely used; however, unlike the rotating systems in medical CT, stationary X-ray systems are more common, and publicly available reconstructed data are limited. This makes it challenging to obtain large-scale 3D labeled data and voxel representations essential for training. To address these limitations, our study presents a calibration and reconstruction method using an unaligned sparse multi-view X-ray baggage dataset, which has extensive 2D labeling. Our approach integrates multi-spectral neural attenuation field reconstruction with Linear pushbroom (LPB) camera model pose optimization, enhancing rendering consistency for novel views through color coding network. Our method aims to improve generalization within the security baggage inspection domain, where generalization is particularly challenging. 
\end{abstract}


\section{Introduction}

\hspace{1em} In security baggage screening, 2D X-ray systems have been widely employed as essential tools for security applications. Leveraging the high penetrative power of X-rays, these systems effectively identify prohibited items in carry-on luggage. If 3D reconstruction is not performed, they allow real-time screening of baggage. Among commonly used systems are those employing two stationary sources and detectors positioned at 90-degree angles, as shown in Figure 1(a), often utilizing multi-energy techniques to produce color-mapped images \cite{Fornaro2011}.
Multi-energy X-ray imaging technology uses multiple energy levels to generate X-ray images, facilitating a deeper understanding of the object's density and effective atomic number \cite{10005308}. By assigning distinct colors to material properties such as organic and inorganic substances, these systems provide enriched visual information, aiding screening officers in distinguishing and identifying various objects with greater efficiency.

\hspace{1em}Through 3D reconstruction, the interpretability of X-ray images can be significantly enhanced. Unlike 2D images, which are limited to specific viewing angles, 3D rendering enables arbitrary-angle visualization, allowing for more accurate assessments. However, achieving CT reconstruction typically requires X-ray images captured from hundreds of angles using rotational gantry systems. Moreover, traditional reconstruction methods are computationally intensive and time-consuming, making them unsuitable for real-time applications.

\hspace{1em}To address these challenges, recent studies have explored generalizable 3D reconstruction using deep learning-based generative models such as GANs and diffusion models. Leveraging the expressive power of these generative models, it becomes feasible to train systems capable of producing 3D representations using only a limited number of X-ray images\cite{CoronaFigueroa2022MedNeRFMN, Liu_2023_ICCV, coronafigueroaa23unaligned}. Despite these advances, most generative models require a substantial amount of ground truth-level 3D data for effective training, while achieving high-quality CT reconstruction with sparse-view X-ray images remains a challenging task. 

\begin{figure}[H]
    \centering
    \includegraphics[width=1.0\textwidth]{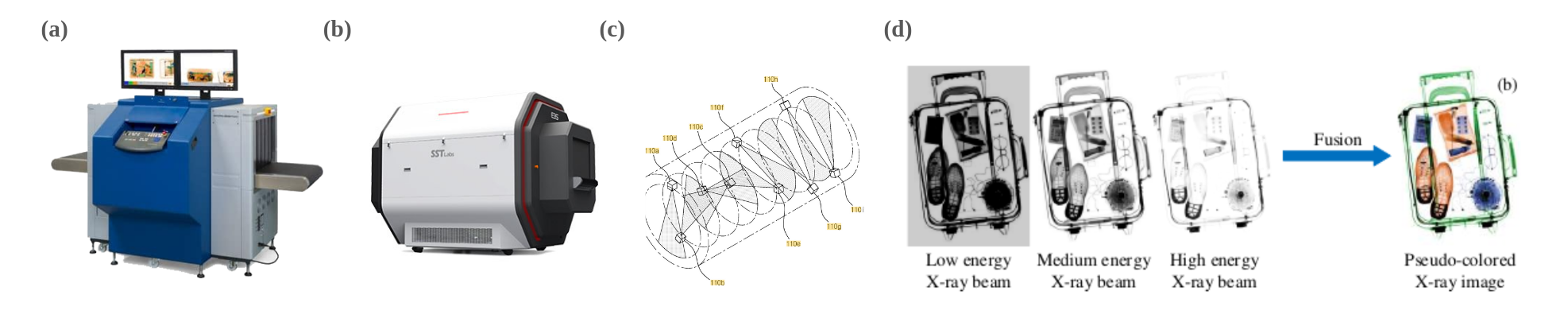}
    \captionsetup{font=scriptsize}
    \caption{(a) Dual view X-ray imaging system by Smith Detection\cite{smithsdetection}, (b) The 9-view X-ray system employed to acquire the multi-view security baggage dataset by SSTlab\cite{sstlabs}, (d) A multi-energy X-ray image and a color coding function utilized to generate a color image. Image adapted from \cite{10005308}}
    \label{fig:image}
\end{figure}

\hspace{1em} Recent studies utilizing Implicit Neural Representations (INR) have shown remarkable success in the field of CT imaging \cite{molaei2023implicit}. For example, they have been effectively applied to tasks such as image reconstruction, segmentation, registration, and novel view synthesis. The exceptional learning capabilities and representational power of Multi-Layer Perceptrons (MLPs) have demonstrated superior performance compared to traditional methods.
Inspired by these advancements, we propose a 3D reconstruction pipeline leveraging INR approaches for sparse-view X-ray baggage datasets. Unlike most existing sparse-view tomographic reconstruction method that assume precise calibration, our method incorporates pose refinement starting from noisy initial estimations.

To summarize, the key contributions of our proposed method are as follows:

\begin{itemize}
\item We propose a CT reconstruction and pose refinement method tailored for sparse-view X-ray baggage datasets that lack precise calibration information. Our method enables the generation of 3D data and labels directly from 2D multi-view images and their annotations.
\item We design a multi-spectral neural attenuation field and a color-coding network, enabling reconstruction from color-mapped RGB X-ray images without requiring raw data of multi-energy X-ray images.
\end{itemize}

\section{Related works}

\textbf{Linear pushbroom camera model: } The linear pushbroom (LPB) camera model is a 1D imaging system widely used in applications such as satellite images and hyperspectral imaging. Unlike the traditional pinhole camera model, the LPB model exhibits perspective properties along one axis and orthographic properties along the other. Theoretical properties of the LPB model can be found in \cite{Gupta1997}. A planar checkerboard-based calibration method, such as the Zhang method, has been investigated in \cite{8296238}. The LPB camera combines perspective and affine transformations, allowing it to accommodate the linear scanning motion typical of platforms such as satellites or drones. This makes it particularly suitable for capturing large-scale area coverage.
In fan-beam CT systems, such as the one used in our dataset, a stationary fan beam scans objects transported along a conveyor belt, adhering to the principles of the LPB model.

\textbf{Joint optimization of NeRF and camera pose refinement: }NeRF demonstrates exceptional performance in novel view synthesis but requires highly accurate camera poses. These poses are typically obtained through off-the-shelf SfM pipelines like COLMAP \cite{7780814}. However, in sparse-view settings, achieving precise camera pose estimation can be challenging. Several approaches aim to reduce NeRF's dependency on accurate camera poses. SCNeRF \cite{SCNeRF2021} introduces a self-calibration algorithm that jointly learns geometry and camera parameters using projected ray distances. BARF \cite{lin2021barf} employs a coarse-to-fine registration strategy to jointly optimize the radiance field and the camera parameters from initially noisy poses. SPARF \cite{sparf2023} proposes a multi-view correspondence loss with a pre-trained matching model and a depth consistency loss using novel view ray sampling in training. GARF \cite{chng2022gaussian} leverages Gaussian activations to propose a positional embedding-free NeRF framework for pose estimation.

\textbf{CT reconstruction with Implicit Neural Representation: } In traditional CT, analytical and iterative methods such as FDK \cite{Feldkamp:84}, MBIR \cite{Willemink2013}, SART \cite{ANDERSEN198481}, ASD-POCS \cite{Sidky2008} have been widely used. These approaches often require hundreds of X-ray projections to achieve high-quality reconstructions and are prone to artifacts under sparse-view conditions. With advancements in deep learning, reconstruction methods leveraging Implicit Neural Representations (INR) have shown promising results in sparse-view settings. These architectures learn functions that map input coordinates to scene properties, enabling impressive 3D scene reconstructions. For instance, in NAF \cite{Zha2022}, attenuation coefficients are used to synthesize projections by simulating the attenuation of incident X-rays based on the predicted coefficients. SAX-NeRF \cite{sax_nerf} introduces a Lineformer architecture based on Transformers, achieving superior reconstruction performance in handling various and complex structures. NeAT \cite{10.1145/3528223.3530121} proposes a hierarchical neural rendering pipeline leveraging explicit octree representation, enhancing scalability and efficiency. While these methods primarily utilize single-energy X-ray images, we extend this framework by employing multi-energy X-ray data. Specifically, we propose an INR-based approach to effectively reconstruct from color-mapped RGB X-ray images, leveraging the additional spectral information encoded in multi-spectral imaging.

\begin{figure}[H]
    \centering
    \includegraphics[width=0.7\textwidth]{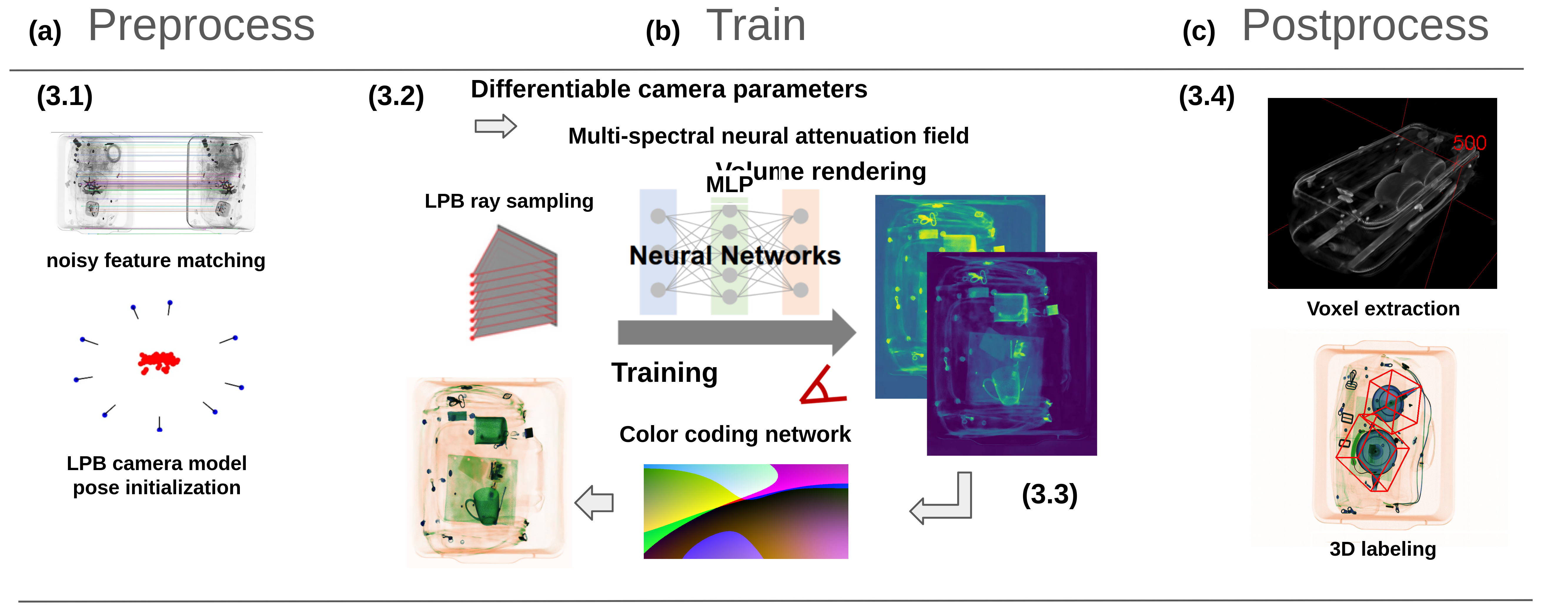}
    \captionsetup{font=scriptsize}
    \caption{Our overall pipeline: (a) LPB camera pose initialization from feature matching. (b) Training stage, where we jointly train multi-spectral neural attenuation field and color coding network. (c) In the post-processing stage, cuboid annotations are generated from 2D bounding box labels. }
    \label{fig:method}
\end{figure}

\section{Method}

\subsection{LPB camera model and pose initialization}

X-ray images were captured using a fan-beam setup, which required the use of an LPB camera model \cite{Gupta1997}. Unlike the conventional pinhole camera model, the LPB camera model is defined by a combination of a perspective and an affine camera model. The intrinsic matrix and the projection is described in Eq. (1).

\begin{equation}
    \begin{bmatrix} u \\ v \\  1 \end{bmatrix} \sim
    \begin{bmatrix} fX + C_xZ \\ sYZ \\  Z \end{bmatrix} = 
    \begin{bmatrix}
        f & 0& C_x \\ 
        0& s& 0 \\  
        0& 0& 1
    \end{bmatrix} \begin{bmatrix} X \\ YZ \\  Z \end{bmatrix} 
\end{equation}

The initial extrinsic parameter was estimated through a bundle adjustment process based on feature matching, similar to traditional structure-from-motion (SfM) pipelines. The LP fundamental matrix \cite{Gupta1997} was computed to perform the image matching under the epipolar constraints. Subsequently, 3D reconstructed points and reprojection errors were calculated to optimize the camera pose. Unlike traditional SfM pipelines, in X-ray imaging, the penetration of X-rays through objects introduces challenges, as feature matching alone tends to be highly imprecise, leading to noisy camera pose estimations. However, since the reconstruction process, discussed later, includes pose refinement, the initial pose estimation step can be simplified or omitted.

\subsection{Multi-spectral neural attenuation field}

Following the methodology of the NAF \cite{Zha2022}, we predict the attenuation coefficient from sampled points. According to Beer’s law, the intensity of X-rays transmitted through a material decreases based on the exponential integration of the attenuation coefficient. The X-rays are sampled along the ray direction $\textbf{r}(t;\mathcal{P}) = \textbf{o} + t\textbf{d}$
based on the LPB camera parameters $\mathcal{P}$, and the intensity at a pixel \textbf{p} in the projected X-ray image is expressed as:

\begin{equation}
    \hat{I}_{}(\textbf{r};\mathcal{P}) = \hat{I}_{o}  \exp (-\int_{t_{\textit{n}}}^{t_{\textit{f}}} \mu_{}({\textbf{r}(t;\mathcal{P})}) dt), 
\end{equation}

Here, $t_{\textit{n}}$ and $t_{\textit{f}}$ represent the ray parameters corresponding to the entry and exit points of the ray within the scene.
Eq. (2) is usually performed in logarithmic space, by substituting $\log(\hat{I}) = I$,

\begin{equation}
    I_{o}(\textbf{p}) - {I}(\textbf{p}) =  \int_{t_{\textit{n}}}^{t_{\textit{f}}} \mu_{}({\textbf{r}(t;\mathcal{P})}) dt),
\end{equation} 

By discretizing Eq. (3), the integral is approximated as:

\begin{equation}
    I_{o}(\textbf{p}) - {I}(\textbf{p}) =  \sum \mu_i\delta_i, 
\end{equation}

where $\delta_i$ represents the distance between the i-th sample and its neighboring sample. We modeled the camera parameters $\mathcal{P}$ with 10 degrees of freedom: intrinsic parameters $f$, $C_x$, two distortion parameters, and 6 degrees of freedom for extrinsic parameters $R|T$. Since the conveyor belt operates at a constant speed, s is shared. The points sampled via the LPB camera parameters are used to predict the attenuation coefficient through positional encoding and a multi-layer perceptron (MLP). However, unlike prior studies, we predict n-dimensional attenuation coefficients, which will be addressed further section.

\subsection{Color coding network}
The dataset employed in this study comprises RGB images generated by color coding of multi-energy CT data. Multi-energy CT facilitates material differentiation by analyzing the attenuation of X-ray spectra, enabling the examination of both Compton scattering and photoelectric effects \cite{Alvarez1976}. The attenuation coefficient of a material is expressed in Eq. (5), which represents the contributions of the photoelectric effect $\mu_{photo}$  and Compton scattering $\mu_{comp}$ as energy-dependent functions weighted by $\alpha$ coefficients.

\begin{equation}
\mu(E) = \mu_{comp}(E) + \mu_{photo}(E) = \alpha_{comp} f_{comp}(E) + \alpha_{photo} f_{photo}(E),
\end{equation}

By leveraging multi-energy X-rays, attenuation coefficients can be decomposed across multiple energy levels.  The aplha coefficient is used through an RGB mapping function, allowing material properties to be more intuitively interpreted by humans.
Motivated by this concept, a color-coding network was designed to extend the output of the attenuation field to n-dimensions, which are then processed through additional network layers to generate RGB images. The MLP estimates the multi-spectral attenuation coefficients, which are then passed to the color-coding network. The color-coding network $F_c$ encodes accumulated outputs of the multi-spectral attenuation field.  Predicted color at pixel \textbf{p} is represented as:

\begin{equation}
\hat{I}_{pred}(\textbf{p}) = F_{c}(\hat{I}_{N,o} \exp (-\int_{t_{\textit{n}}}^{t_{\textit{f}}} \mu_{}({\textbf{r}(t;\mathcal{P})}) dt);\theta_{c})
\end{equation}

The training objective is to minimize the squared error $\mathcal{L}$ between ground truth pixel value and predicted one. We jointly update camera parameter $\mathcal{P}$, MLP of multi-spectral neural attenuation field and color coding network. 

\begin{equation}
\mathcal{L(\mathcal{P},\theta_{\textit{m}},\theta_{\textit{{c}}})} = \sum ||I_{gt}{(\textbf{p})}-I_{pred}{(\textbf{p})} ||^2_2
\end{equation}

Through these processes, the color-coding network implicitly learns a mapping function from multispectral data to RGB representations. This approach effectively mitigates artifacts that arise from incorrect mappings between RGB colors and attenuation coefficients.

\subsection{Generating cuboid labels from 2D bounding box annotations}

The cuboid labeling data is post-processed after the reconstruction process is completed. In this step, simple yet efficient methods the visual hull \cite{273735}, rotating calipers \cite{Toussaint1983SolvingGP} and region-growing based segmentation method propsed in \cite{Wiley2012} are used. For the n-th voxel coordinates and the K-th image, visual hull is performed.

\begin{algorithm}[H]
\SetAlgoLined
\KwIn{$N^3$ voxels, $K$ images, projection function $\Pi^k$}
\KwOut{Visual hull of the object}

\For{$n = 1$ to $N^3$}{
    \For{$k = 1$ to $K$}{
        Project point $p$ onto the $k^{th}$ image plane using $\Pi^k$\;
        \If{$\Pi^k(p) \in \mathbb{R}^2$ lies outside the bounding box}{
            Exclude point $p$ in the visual hull\;
        }
    }
}
\caption{Visual Hull}
\end{algorithm}

Since the bounding box annotations in the 2D images lack rotation information, the resulting visual hull does not tightly enclose the object. To address this, object segments are subsampled through the process described in \cite{Wiley2012}. Subsequently, the rotating calipers \cite{Toussaint1983SolvingGP} algorithm is applied to generate cuboid labels that tightly and compactly enclose the object.

\section{Experimental results}

\subsection{Datasets}

The dataset used in this study was constructed as part of the data development initiative by Korea's National Information Society Agency \cite{NIA}. The X-ray system is equipped with nine independent fan-beam sources and detectors. Images are captured as objects are transported on a conveyor belt. Each detector is positioned at an angle of approximately 40 degrees. However, as the system does not operate in a rotational format around an axis, precise calibration information is not provided. The images are processed into RGB color maps derived from dual-layer detectors \cite{Fornaro2011} using two different energy levels of X-rays, while raw data is not provided. The dataset consists of 541,260 2D images, categorized into 317 classes including "Prohibited Carry-On Items," "Data Storage Media," and "General Items." Each object in the dataset is annotated with 2D bounding boxes and segmentation labels. The ground-truth level reconstructed voxel data is not provided.

\begin{figure}[H]
    \centering
    \includegraphics[width=1.0\textwidth,height=0.37\textwidth]{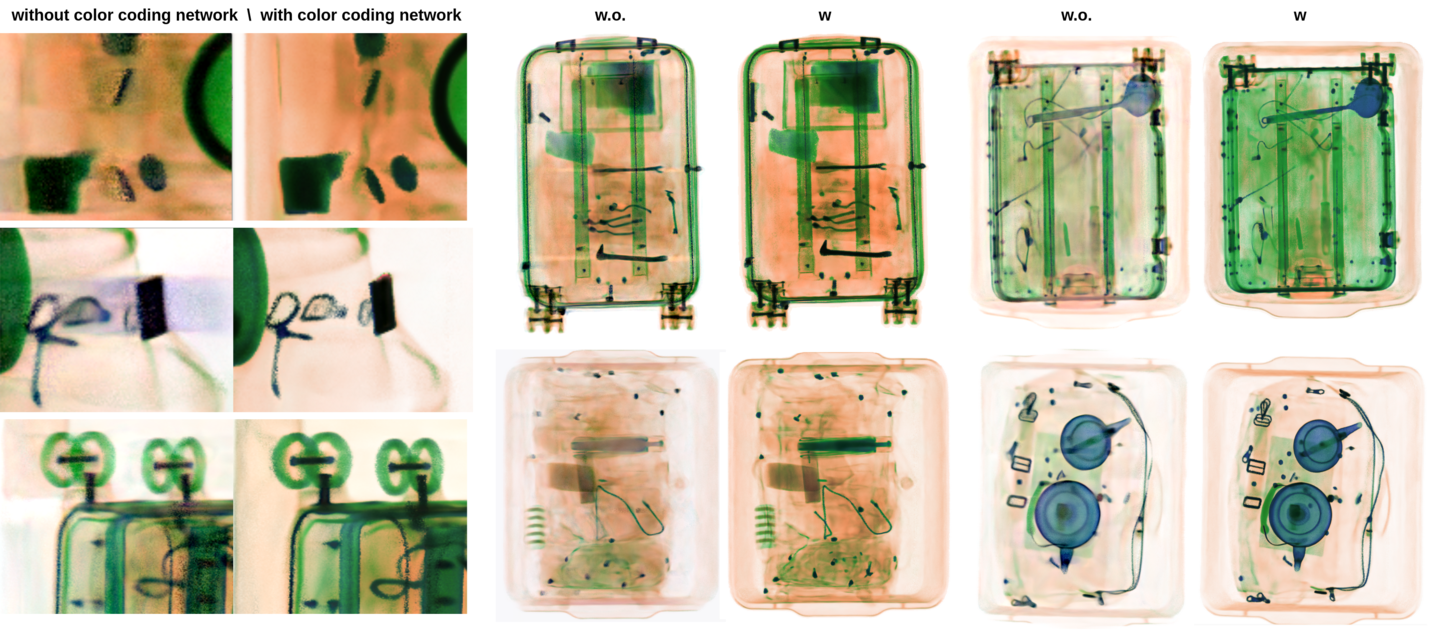}
    \captionsetup{font=scriptsize}
    \caption{Comparison of images without and with the use of the color coding network. }
    \label{fig:figure1}
\end{figure}

\begin{figure}[H]
    \centering
    \includegraphics[width=1.0\textwidth]{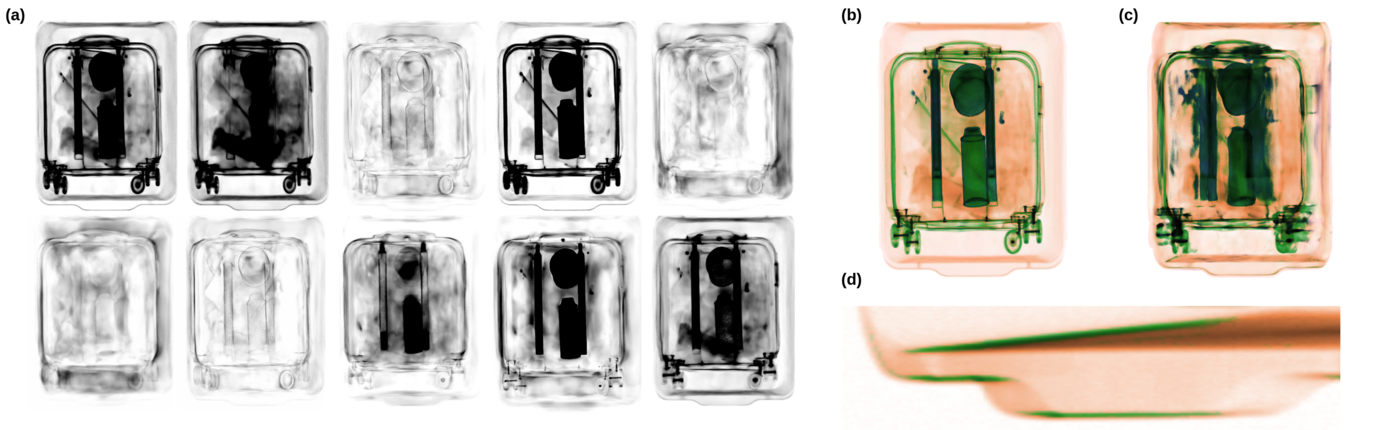}
    \captionsetup{font=scriptsize}
    \caption{(a) Visualization of the neural attenuation field at n=10, (b) rendered image trained at n=10, (c) rendered image trained at n=1, (d) Material-color mismatching observed in the ground truth image due to ambiguity arising from the viewing angle. }
    \label{fig:figure2}
\end{figure}

\subsection{Method analysis}

\textbf{Impact of color coding network: } Our proposed network employs a color coding network to implicitly learn the n-dimensional attenuation coefficients and the corresponding color mapping function. This design aligns with the image processing workflow of multi-energy X-ray datasets, in contrast to mono-energy X-ray images. As illustrated in Figure 3, our approach yields reduced artifacts and sharper reconstructed results. Given the unique characteristics of security baggage datasets—unlike medical CT scans—they often include metallic components that can introduce significant metallic artifacts \cite{6922068} with traditional reconstruction algorithm. The presence of high-density objects within the scan field of view leads to the generation of artifacts \cite{Mouton2013AnES}. We observed noticeable color smearing effects near the deep blue objects during the reconstruction process. Due to the fan-beam configuration, these smearing effects are observed along horizontal lines. It can be inferred that such phenomena arise from the non-linearity between attenuation coefficients and the imaged colors. 

\hspace{1em} Notably, our color coding network effectively mitigates these issues. Figure 4(a) provides a visualization of the attenuation fields learned using a 10-dimensional multi-spectral neural attenuation field. It highlights how features within the attenuation field are optimized to make the classification of various material properties more accessible for the color coding network. This indicates the network's ability to adaptively distinguish material-specific features, resulting in enhanced performance in multi-energy imaging tasks. Theoretically, since our data consists of color-mapped images derived from two energy levels, the dimensionality of the multi-spectral neural attenuation field requires at least two dimensions. Figure 4(c) shows the result obtained when the color coding network is trained with only one dimension, demonstrating inferior performance compared to the case where the color coding network is not used. Figure 4(d) illustrates material-color mismatching present in the ground truth image, a characteristic our network inherently mimics. The material of the same type is mapped to different colors due to the viewing angle, and the trained network exhibits the same characteristic.

\begin{table}[h!]
\centering
\begin{tabular}{lccc}
\toprule
\textbf{Method} & \textbf{PSNR↑} & \textbf{SSIM↑} & \textbf{LPIPS↓} \\
\midrule
w.o.  & 20.93 & 0.87 & 0.30 \\
n = 1 & 12.98 & 0.74 & 0.37 \\
n = 2 & 21.41 & 0.92 & 0.23 \\
n = 3 & 22.48 & 0.90 & 0.20 \\
n = 5 & 22.31 & 0.89 & 0.19 \\
n = 10 & 20.41 & 0.88 & 0.21 \\
\bottomrule
\end{tabular}
\vspace{1ex}
\captionsetup{font=scriptsize}
\caption{Comparison of evaluation metrics across different methods. n indicates the number of dimensions in the color coding network.}
\label{tab:evaluation_metrics}
\end{table}

\textbf{Pose refinement and positional encoding: } We tested two types of positional encoders: frequency encoder \cite{tancik2020fourfeat} and hash encoder\cite{mueller2022instant}. With our naive implementation, hash encoder showed faster convergence; however, in novel view rendering, the frequency encoder produced qualitatively better results. This is likely because, in the model architecture that simultaneously performs pose refinement and scene reconstruction using only a training signal, high-frequency signals are essential for pose optimization. A deterministic mapping of input 3D coordinates to higher dimensions through different sinusoidal frequency bases aids in this optimization. While the hash encoder also holds potential for better results, improvements through the introduction of a regularizer that can provide signals necessary for pose optimization are anticipated.

\section{Conclusion}

This paper introduces a CT reconstruction method for aligned sparse-view color-coded X-ray images. By integrating the LPB camera model, our approach incorporates pose refinement into the reconstruction process, addressing the challenges of noisy initial poses. The introduction of the color coding network enables the implicit learning of a color mapping function during reconstruction, significantly enhancing the quality of novel view synthesis, even without access to raw multi-energy data. Looking ahead, we plan to incorporate Gaussian splatting\cite{kerbl3Dgaussians}, aiming to achieve faster rendering speeds suitable for real-time applications.

\section*{Acknowledgments}

\bibliographystyle{unsrt}  
\bibliography{references}  

\nocite{kerbl3Dgaussians}
\nocite{mueller2022instant}
\nocite{tancik2020fourfeat} 
\nocite{mildenhall2020nerf}


\end{document}